\documentstyle[12pt,amsbsy]{article}
\begin{document} 
\newcommand{\om}{{\vec \omega}}
\newcommand{\B}{{\vec B}}
\newcommand{\E}{{\vec E}}
\newcommand{\V}{{\vec v}}

\begin{titlepage}

\vspace{2cm}

\begin{center} 
{\large \bf  Feasibility of search\\ for nuclear electric dipole moments\\
at ion storage rings\\}
\end{center}
\vspace{1cm}

\begin{center} 
I.B. Khriplovich\footnote{E-mail address: khriplovich@inp.nsk.su}\\
\end{center}

\begin{center}
Budker Institute of Nuclear Physics,\\
630090 Novosibirsk, Russia,\\
and Novosibirsk University, Novosibirsk, Russia
\end{center}

\vspace{2cm}

\begin{abstract}
The sensitivity much better than $10^{-24}\;e$ cm may be expected in the 
searches for electric dipole moments (EDM) of $\beta$-active nuclei at 
ion storage rings. It would be a serious progress in studies of the 
CP-violation problem.
\end{abstract}

\vspace{4.5cm}

\end{titlepage}

{\bf 1.} Only upper limits have been obtained up to now in the 
searches 
for the electric dipole moments of the neutron, atoms and molecules. 
But these limits are a valuable contribution to elementary particle 
physics, having strongly constrained theoretical models of CP
violation. 

Let us note that the bounds on CP violation in nuclei derived from 
atomic 
experiments are at least as informative as direct investigations of the 
neutron EDM. New approaches may arise here due to the progress in the
accelerator technique. An experiment was recently proposed to 
search for the muon EDM with the sensitivity of 
$10^{-24}\;e$ cm~\cite{pro}. The 
intention is to use the existing muon $g-2$ ring. The muons in it 
have natural longitudinal polarization. An additional spin precession 
due to the EDM interaction with external field will be monitored by 
counting the decay electrons, their momentum being correlated with the 
muon spin due to parity nonconservation in the muon decay.

Just for the coherence of presentation, let us write down few 
formulae from~\cite{pro} with simple explanations. The frequency 
${\om}_m$ of the 
spin precession in external magnetic and electric fields, ${\B}$ and 
${\E}$, is
(see, e.g., book~\cite{blp}, \S 41)
\begin{equation} \label{bmt}
{\om}_m = - {e \over m} \left\{\left(a+{1 \over \gamma}\right)\,{\B} 
- 
a\,{\gamma \over \gamma +1}\,{\V}\,({\V}{\B}) -
\left(a + {1 \over \gamma +1}\right) {\V} \times {\E} \right\}. 
\end{equation}
Here the anomalous magnetic moment $a$ is related to the $g$-factor 
as follows:
$a=g/2-1$, for muon $a=\alpha/2\pi$; ${\V}$ is the particle velocity;
$\gamma=1/\sqrt{1 - v^2}$. The units are used where $\hbar=1$, 
$c=1$. We supply 
here ${\om}$ by subscript $m$ to indicate that formula (\ref{bmt}) 
describes 
the precession due to the magnetic moment (combined with the Thomas 
effect). As to the 
precession induced by the EDM, its frequency is
\begin{equation} \label{bet}
{\om}_e = - {e \over m}\,\eta\,\left\{{\E} - 
{\gamma \over \gamma +1}\,{\V}\,({\V}{\E}) + {\V} \times {\B} 
\right\}. 
\end{equation}
The dimensionless constant $\eta$ is related to the EDM $d$ as 
follows:
\[ d = {e \over 2m}\,\eta \]
(here $\eta$ is 2 times smaller than in~\cite{pro}).
Formula (\ref{bet}) can be obtained from the terms proportional to 
the anomalous
magnetic moment in (\ref{bmt}) by substituting $a\rightarrow \eta$ 
and changing
to dual fields: ${\B}\rightarrow {\E}$, ${\E}\rightarrow -{\B}$.

However, what is of interest to this muon EDM experiment, is not the 
frequency of the 
spin precession with respect to the laboratory frame, i.e., not 
the sum 
of expressions (\ref{bmt}) and (\ref{bet}). What we need is the 
frequency of the spin precession 
with respect to the muon momentum. The precession 
frequency of the momentum 
itself ${\om}_p$ can be derived easily from the well-known 
expression for acceleration
in external fields (see, e.g., book~\cite{ll}, problem to \S 18):
\begin{equation} \label{ac}
\dot{{\V}} = {e \over m\gamma} \left\{{\V} \times {\B} + {\E} - 
{\V}\,({\V}{\E}) \right\}. 
\end{equation}
The acceleration component transverse to the velocity is
\begin{equation} \label{act}
\dot{{\V}}_t = {e \over m\gamma} \left\{{\V} \times {\B} - 
{\V}\times[{\V}\times{\E}] \right\}, 
\end{equation}
which corresponds to the precession frequency of momentum (or
velocity)
\begin{equation} \label{omp}
{\om}_p = - {e \over m} \left\{{1 \over \gamma}\,{\B} -
{\gamma \over \gamma^2 - 1}\,{\V} \times {\E} \right\}. 
\end{equation}
Thus, the frequency of the spin precession with respect to the 
momentum is
\[
{\om} = {\om}_m + {\om}_e - {\om}_p = - {e \over m} \left\{a {\B} - 
a\,{\gamma \over \gamma +1}\,{\V}\,({\V}{\B}) -
\left(a - {1 \over \gamma^2 - 1}\right) {\V} \times {\E} \right.\] 
\begin{equation} \label{sp} 
\left.+ \eta \left[{\E} - {\gamma \over \gamma +1}\,{\V}\,({\V}{\E})
+{\V} \times {\B}\right]\right\}. 
\end{equation}
This expression simplifies in the obvious way at 
$({\V}{\B})=({\V}{\E})=0$.
Just this case is considered below.

It is proposed in~\cite{pro} to compensate for the precession in 
the vertical magnetic field ${\B}$ by the precession in a radial 
electric field ${\E}$, i.e., to choose ${\E}$ in such a way that
\[a{\B}\,-\,\left(a - 
{1 \over \gamma^2 - 1 }\right) {\V} \times {\E}\,=\,0\,. \]
In fact, electric fields in a storage ring are much smaller than 
magnetic ones, and therefore can be 
neglected in the EDM term. So, with the mentioned compensation, 
the spin precession with respect to momentum is due only to the 
EDM interaction with the vertical magnetic field:
\begin{equation} \label{edm}
{\om} ={\om}_e = - {e \over m}\,\eta\, {\V} \times {\B}. 
\end{equation}
In this way the muon spin acquires a vertical component which 
linearly grows with time. The P-odd 
correlation of the decay electron momentum with the muon spin 
leads to the 
difference between the number of electrons registered above and 
below the orbit plane.

The statement in~\cite{pro} on the feasibility of this experiment   
is: "We are confident that we can improve by six 
orders of magnitude the current sensitivity to the muon EDM, both in 
statistics and systematics, bringing it down to $10^{-24}\;e$ cm."

\bigskip
\bigskip
 
{\bf 2.}  Our point is that in the same way one can search for an EDM of 
a polarized 
$\beta$-active nucleus in a storage ring. In this case as well,
the precession of nuclear spin due to the EDM interaction can be 
monitored by the direction of the $\beta$-electron momentum. 

$\beta$-active nuclei have serious advantages as compared 
to muon. 

The life-time of a $\beta$-active nucleus can exceed by many orders 
of magnitude that of a muon. The characteristic depolarization time of 
the ion beam is also much larger than $10^{-6}$ s, the muon life-time. 
Correspondingly, the angle of the rotation of nuclear spin, which is 
due to the EDM interaction and which accumulates with time, may be also
by orders of magnitude larger than that of a muon. By the same reason 
of the larger life-time, the quality of an ion beam can be made much 
better than that of a muon beam.

Then, the typical nuclear magnetic moment is by an order 
of magnitude smaller than that of a muon. Accordingly, smaller are 
various spurious effects due to the interaction of a magnetic moment 
with external fields.

However, necessary conditions here are also quite serious. 

First of all, there should be an appreciable P-odd correlation in 
the $\beta$-decay between the electron momentum and spin $J$ of the 
decaying nucleus.

Then, to make realistic the mentioned compensation of the 
EDM-independent spin
precession by a relatively small electric field, the effective 
$g$-factor should be 
close to 2 (as this is the case for the muon). Let us consider this 
condition for a 
nucleus in more detail. The nuclear magnetic moment
\[ {e \over 2m_p}\,\mu, \]
when expressed through the total nuclear charge $Ze$ and mass 
$Am_p$, can be rewritten as
\[ {Ze \over 2Am_p}\,{A \over Z}\,\mu. \]
The effective $g$-factor and effective anomalous magnetic moment 
are now
\[ g = {A \over Z}\,{\mu \over J}\;\;\; \mbox{and}\;\;\; 
  a = {g \over 2} - 1 = {A \over Z}\,{\mu \over 2J} - 1, \]
respectively. Therefore, the condition discussed is
\begin{equation} \label{con} 
{A \over 2 Z}\,{\mu \over J} \approx 1,\;\;\; \mbox{or}\;\;\; 
 \mu \approx {2 Z \over A} J \approx 0.8 J. 
\end{equation}
Fine-tuning to this condition is possible in many cases by taking, 
instead of a 
bare nucleus, an ion with closed electron shells. Then the necessary 
condition (\ref{con}) softens to
\begin{equation} \label{ft}
\mu \leq {2 Z \over A} J \leq 0.8 J. 
\end{equation}
An even number of electrons in an ion does not
guarantee by itself that the total electron angular momentum, and 
therefore the total electron magnetic moment vanish. Its vanishing
can be demonstrated at least at the following numbers of electrons 
\[ Z-z = 2,\;4,\;6,\;10,\;12,\;14,\;18, \]
since up to 
\[ 1s^2\, 2s^2\, 2p^6\, 3s^2\, 3p^6 \] 
the filling of atomic shells certainly follows the hydrogen pattern. 
On the other hand, at $Z-z = 8,\;16\,,$ for the ground state
configurations of the type $p^4$, when the $p$-subshell is filled 
more than by half, the total electron angular momentum $J_e = 2$. 
One should exclude also the case $Z-z = 26$: neither of the electron
configurations conceivable here, 
$\;\;3d^6 4s^2$\nolinebreak  $\;^5D_4,\;\;\;
3d^7 4s$\nolinebreak  $\;^5F_5,\;\;\;3d^8$ \nolinebreak $^3F_4,\;\;$ 
has vanishing angular momentum. In other cases of interest vanishing of 
$J_e$ should be checked experimentally.

One more comment on the ion charge $z$ should be made. The nuclear EDM 
is screened 
by atomic electrons, and the screening is complete for a neutral atom. 
It is 
intuitively clear and can be confirmed by direct calculations that in 
an ion this 
screening is partial only, being proportional to the number of 
electrons $Z-z$. 
So, the smaller is this number, the better for our problem.

Finally, if we take into account the difference $\Delta m = m_n-m_p$ 
between the neutron and proton masses, and the finite mass $m_e$ 
of the 
electron, the expression for
the effective anomalous magnetic moment of an ion, with the nuclear 
charge $Z$ 
and the total ion charge $z$, changes to
\begin{equation}
a = {A\over 2z}\,{\mu \over J}\left(1+{A-Z \over A}\,{\Delta m \over 
m_p}\,+\,
{Z-z \over A}\,{m_e \over m_p}\right) - 1. 
\end{equation}
Typically, due to the correction factor 
\[ \left(1+{A-Z \over A}\,{\Delta m \over m_p}\,+\,
{Z-z \over A}\,{m_e \over m_p}\right)\,,\]
the value of $a$ increases by $\;(8\; -\; 9)\cdot 10^{-4}$.

The ions which look at the moment promising from the point of view 
of the EDM
searches are presented in the Table. The isotope data are taken
from the handbook \cite{tab}. The selection is confined to those
$\beta$-active nuclei for which the positive sign of magnetic moment 
is established: at negative sign the mentioned fine-tuning is 
impossible at all. The next demand was that the value of magnetic 
moment should be known with sufficient precision and should allow 
reasonable fine-tuning. 

The errors in the values of anomalous magnetic moments $a$ presented
in the Table correspond to the experimental errors in values of $\mu$.
But there is an effect not taken into account in the Table: electron 
configurations even with vanishing angular momentum $J_e$ produce a 
diamagnetic screening of nuclear magnetic moments. The
screening should be very small in case of closed electron shells, 
$s^2$, $p^6$. But it is nonnegligible in the configurations of the 
type $p^2$, i.e., at $Z-z = 6,\;14\,$. The relative magnitude of the
diamagnetic effect can be estimated here roughly as $1/Z$. The 
diamagnetic correction is truly large for $^{24}_{11}$Na,
changing the $a$-value from 0.015 presented in the Table to about
$\;\;$- 0.1. The exact value of $a$ for $^{24}_{11}$Na at $z=5$ 
demands rather serious atomic calculations and/or experimental 
measurements. But most probably it will stay relatively large. This is
quite unfortunate since just $^{24}_{11}$Na$^{+5}$ seems to be a good 
object for the discussed experiments, being available in very large 
quantities.

We have excluded from the Table isotopes with too short and too 
long life-time $t_{1/2}$ (for our purpose approximate values of 
$t_{1/2}$ are sufficient, so we present them with 2 digits only). In 
this respect at least $^{137}_{\;\;55}$Cs looks already suspicious. By 
the way, few examples of $\beta$-decaying excited states satisfy the 
above criteria, these isotopes are marked in the Table by *. The 
optimum values of 
life-times depend on too many experimental details and therefore cannot 
be indicated in general form. 

The value $J=1/2$ for nuclear spin would allow to avoid the background
due to the quadrupole interaction with external fields. Unfortunately, 
we could not find spin $J=1/2$ isotopes satisfying our criteria. 
Simple estimates demonstrate however, that at reasonable parameters of 
a storage ring even for relatively large nuclear quadrupole moments 
$Q \sim 1$ barn this background is not dangerous at the EDM sensitivity
as high as $10^{-26}\;e$ cm. The values of $Q$ (where known) are also 
presented in the Table with 2 digits only.
 
All isotopes presented in the Table are $\beta^-$-active (their
$\beta^-$ branchings are indicated in the last column). Fortunately,
many of them have allowed pure Gamow -- Teller transitions 
($|\Delta J| = 1$) where the magnitude of the needed correlation 
between the electron momentum and the initial spin is on the order of 
unity. Few isotopes in
the Table have allowed mixed $\beta^-$-transitions ($|\Delta J| = 0$). 
Here the magnitude of the asymmetry we need may change essentially 
from nucleus to
nucleus. For instance, for the cases presented in book \cite{sch} 
(they do not
enter the Table since none of them fits our criteria) this asymmetry 
varies from 
0.016 to 0.33. Obviously, for the allowed mixed transitions, as well as 
for
forbidden transitions which are also represented in the Table, the 
values of the
discussed asymmetry should be found experimentally. 

Perhaps, from the point of view of registration, it would be tempting 
to have isotopes with positron $\beta$-decay. Unfortunately, for two 
isotopes potentially useful in this respect, $^{70}_{33}$As and 
$^{71}_{33}$As, the cited 8th edition of handbook \cite{tab} contains 
no quantitative data on their $\beta^+$ branchings (though these data      
were present in the 7th edition).

\bigskip
\bigskip

{\bf 3.} But how significant would be the discussed experiments with
$\beta$-active nuclei for elementary particle physics? 

The typical value of a nuclear EDM, as induced by CP-odd nuclear 
forces, is roughly independent of $A$ and $Z$ and can be estimated 
as~\cite{sfk}
\begin{equation}
d_N \sim 10^{-21} \xi, 
\end{equation}
where $\xi$ is a dimensionless parameter measuring these forces in 
the units of the Fermi weak interaction constant $G$.
The best upper limit on $\xi$ was obtained in an atomic 
experiment~\cite{lam}:
\begin{equation}
\xi < 2\cdot 10^{-3}. 
\end{equation}
This limit was demonstrated~\cite{kk} to be at least as significant for
elementary particle physics as the best upper limit on the neutron 
EDM~\cite{ill,spb}:
\begin{equation}
d_n < 10^{-25}\;e\;\mbox{cm}. 
\end{equation}
(Detailed discussion of these problems can be found also in 
book~\cite{khr}.)

So, even at the same sensitivity $10^{-24}\;e$ cm, as discussed 
in~\cite{pro}
for muons, the experiments with $\beta$-active nuclei would compete 
with the best EDM studies. Certainly, progress in this direction well 
deserves serious efforts. 

\begin{center}
***
\end{center}

I highly appreciate numerous helpful discussions with A.N. Skrinsky
and Y.K. Semertzidis. I am grateful also to J. Behr, J. Deutsch,
F. Farley, S. Freedman, P. Herczeg, W. Morse and Yu.F. Orlov for
useful comments and criticisms. Part of this work was done during my
stay at the International Center for Relativistic Astrophysics, Rome,
and at the Theoretical Physics Institute, University of Minnesota, I
greatly appreciate their warm hospitality.
The work was supported also by 
the Russian Foundation for Basic Research through Grant No. 98-02-17797
and by the Federal Program Integration-1998 through Project No. 274. 
 
\newpage

\begin{center}
{\bf Table}
\end{center}

\begin{tabular}[h]{|l|c|c|c|c|c|c|c|}
\hline
         &        &                 &       &          &     &  & \\
Ion & $J^{\pi}\rightarrow J^{\pi\;\prime}$ & $\mu$     & $z$ & 
$a\cdot 10^3$ & $t_{1/2}$ & $Q$ (barn)& branching \\
\hline
 &        &                 &       &          &              &   &  \\
$^{\;\;24}_{\;\;11}$Na  & 4$^+\rightarrow 4^+$ & 1.6903(8) & 5  & 
15.1(0.5)  & 15 h  &   & 99.944\%\\
                &   &           &    &          &         &    & \\
$^{\;\;60}_{\;\;27}$Co & 5$^+\rightarrow 4^+$  & 3.799(8) & 23 & - 
8(2)  & 5.3 y  & 0.44  &99.925\%\\
 &        &                 &       &          &              &   &\\ 
$^{\;\;82}_{\;\;35}$Br      & 5$^-\rightarrow 4^-$   & 1.6270(5) & 13  
& 27.2(0.3) & 35 h & 0.75& 98.5\% \\
                &   &           &    &          &         &    & \\
$^{\;\;93}_{\;\;37}$Rb & 5/2$^-\rightarrow 5/2^+$ & 1.4095(16) & 27 
& - 28.1(1.1) & 5.8 s & 0.18  &43\% \\
                &   &           &    &          &         &    & \\
$^{\;\;94}_{\;\;37}$Rb & 3$^-\rightarrow 3^-$ & 1.4984(18) & 23 & 
21.5(1.2) & 2.7 s & 0.16 &30.6\% \\
                &   &           &    &          &         &    & \\
$^{110}_{\;\;47}$Ag*& 6$^+\rightarrow 5^+$  & 3.607(4) & 33 & 3(1) 
& 250 d & 1.4  &66.8\% \\ 
                &   &           &    &          &         &    & \\
$^{118}_{\;\;49}$In* & 8$^-\rightarrow 7^-$ & 3.321(11) & 25 & - 
19(3) & 8.5 s &  0.44&1.4\% \\ 
&        &                 &       &          &              &   & \\
$^{120}_{\;\;49}$In* & (8$^-)\rightarrow 7^-$ & 3.692(4)  & 27 & 
26(1) & 47 s & 0.53 &84.1\%\\ 
&        &                 &       &          &              &   & \\
$^{121}_{\;\;50}$Sn & 3/2$^+\rightarrow 5/2^+$ & 0.6978(10) & 28 & 
6(1) & 27 h & - 0.02(2) &100\%\\
  &        &                 &       &          &              &   & \\
$^{125}_{\;\;51}$Sb & 7/2$^+\rightarrow 5/2^+$ & 2.630(35) & 47 & 
0$\pm$13 & 2.8 y &  &40.3\%\\  
&        &                 &       &          &              &   & \\
$^{131}_{\;\;53}$I & 7/2$^+\rightarrow 5/2^+$ & 2.742(1)  & 51  & 
7.0(0.4) &  8.0 d  & - 0.40&89.9\%\\
  &        &                 &       &          &              &   & \\
$^{133}_{\;\;53}$I  & 7/2$^+\rightarrow 5/2^+$& 2.856(5)  & 53  & 
25(2)  & 21 h      & - 0.27&83\%\\
  &        &                 &       &          &              &   & \\
$^{133}_{\;\;54}$Xe & 3/2$^+\rightarrow 5/2^+$ & 0.81340(7) & 36  
& 2.58(9)  & 5.2 d & 0.14 &99\%\\
  &        &                 &       &          &              &   & \\
$^{134}_{\;\;55}$Cs & 4$^+\rightarrow 4^+$   & 2.9937(9) & 51  & - 
16.0(0.3)   & 2.0 y     & 0.39&70.11\%\\
 &        &                 &       &          &              &   & \\
$^{136}_{\;\;55}$Cs & 5$^+\rightarrow 6^+$   & 3.711(15) & 51  & - 
9(4)  & 13 d      & 0.22 &70.3\%\\
  &        &             &       &          &              &   & \\                                                                      
$^{137}_{\;\;55}$Cs & 7/2$^+\rightarrow 11/2^-$ & 2.8413(1) & 55  
& 11.9(0.1) & 30 y & 0.051&94.4\%\\  
&        &                 &       &          &              &   & \\
\hline
\end{tabular}

\newpage

\begin{center}
{\bf Table} (continued)
\end{center}

\begin{tabular}[h]{|l|c|c|c|c|c|c|c|}
\hline
   &        &             &       &          &              &   &  \\
Ion & $J^{\pi}\rightarrow J^{\pi\;\prime}$ & $\mu$     & $z$ & 
$a\cdot 10^3$ & $t_{1/2}$ & $Q$ (barn)& branching \\
\hline
&        &                 &       &          &              &   & \\
$^{139}_{\;\;55}$Cs & 7/2$^+\rightarrow 7/2^-$ & 2.696(4) & 53 & 
11(1) & 9.3 m& - 0.075&82\%\\
&        &                 &       &          &              &   & \\
$^{141}_{\;\;55}$Cs & 7/2$^+\rightarrow 7/2^-$ & 2.438(10) & 49 & 
3(4) & 25 s & - 0.36&57\%\\
&        &                 &       &          &              &   & \\
$^{143}_{\;\;55}$Cs & 3/2$^+\rightarrow 5/2^-$ & 0.870(4) & 41& 
12(5) & 1.8 s & 0.47&24\%\\
&        &                 &       &          &              &   & \\
$^{140}_{\;\;57}$La & 3$^-\rightarrow 3^+$ & 0.730(15) & 17 & 
3$\pm$21 & 1.7 d & 0.094&44\%\\
&        &                 &       &          &              &   & \\
$^{160}_{\;\;65}$Tb& 3$^-\rightarrow 2^-$ & 1.790(7) & 47 &16(4) & 
72 d & 3.8&44.9\%\\
&        &                 &       &          &              &   &\\ 
$^{170}_{\;\;69}$Tm& 1$^-\rightarrow 0^+$ & 0.2476(36) & 21 & 
2.2$\pm$14.5 & 129 d & 0.74&99.854\%\\
&        &                 &       &          &              &   &\\ 
$^{177}_{\;\;71}$Lu & 7/2$^+\rightarrow 7/2^-$ & 2.239(11) & 57 & 
- 6(5) & 6.7 d & 3.4&78.6\%\\
&        &                 &       &          &              &   & \\
$^{183}_{\;\;73}$Ta & 7/2$^+\rightarrow 7/2^-$ & (+)2.36(3) & 61 & 
12(13) & 5.1 d &  &92\%\\
&        &                 &       &          &              &   & \\
$^{192}_{\;\;77}$Ir& 4$(^+)\rightarrow 3^+,4^+$ & 1.924(10) & 47 & - 
17(5) & 74 d &2.3&
42\%,54\%\\
  &        &                 &       &          &              &   & \\
$^{196}_{\;\;79}$Au & 2$^-\rightarrow 2^+$   & 0.5906(5) & 29  & - 
1.1(8) & 6.2 d & 0.81 &8\%
\\ &        &                 &       &          &              &   & \\
$^{198}_{\;\;79}$Au & 2$^-\rightarrow 2^+$   & 0.5934(4) & 29  & 
13.9(7) & 2.7 d & 0.68 &98.99\%\\
 &        &                 &       &          &              &   & \\
$^{203}_{\;\;80}$Hg & 5/2$^-\rightarrow 3/2^+$ & 0.84895(13)&34  
& 14.71(15)     & 47 d & 0.34&100\%\\
 &        &                 &       &          &              &   & \\ 
$^{222}_{\;\;87}$Fr & 2$^-\rightarrow 3^-$ & 0.63(1) & 35 & 0$\pm$ 
20 & 14 m & 0.51&55\%\\
         &        &        &       &          &              &   & \\
$^{223}_{\;\;87}$Fr & 3/2$(^-)\rightarrow 3/2^-$ & 1.17(2) & 87 & 
0$\pm$ 20& 22 m & 1.2&67\%\\
 &        &                 &       &          &              &   & \\
$^{224}_{\;\;87}$Fr & 1$(^-)\rightarrow 1^-$ & 0.40(1) & 45 & - 
$3\pm$ 25& 3.3 m & 0.52&42\%\\
&        &                 &       &          &              &   &\\ 
$^{242}_{\;\;95}$Am & 1$^-\rightarrow 0^+,2^+$ & 0.3879(15) &  47 
& - 0.5$\pm$ 3.9& 16 h & - 2.4&37\%,46\% \\
&        &                 &       &          &              &   &\\ 
\hline
\end{tabular}


\newpage

\begin{thebibliography}{99}

\bibitem{pro} Y.K. Semertzidis, in: Proceedings of the Workshop on 
Frontier Tests of Quantum Electrodynamics and Physics of the Vacuum, 
Sandansky, Bulgaria, June 1998, in press.

\bibitem{blp} V.B. Berestetskii, E.M. Lifshitz and L.P. Pitaevskii, 
Quantum Electrodynamics (Pergamon Press, 1994).

\bibitem{ll}  L.D. Landau and E.M. Lifshitz, The Classical Theory of 
Fields (Butterworth-Heinemann, 1995).

\bibitem{tab} R.B. Firestone et al., Table of Isotopes, 8th Edition (John 
Wiley, 1996).

\bibitem{sch} H.F. Schopper, Weak Interactions and Nuclear Beta 
Decay (North Holland, 1966).

\bibitem{sfk} O.P. Sushkov, V.V. Flambaum and I.B. Khriplovich, Zh. 
Eksp. Teor. Fiz. 87 (1984) 1521 [Sov. Phys. JETP 60 (1984) 873].

\bibitem{lam} S.K. Lamoreaux et al., Phys. Rev. Lett. 59 (1987) 2275.

\bibitem{kk} V.M. Khatsymovsky and I.B. Khriplovich, Phys. Lett. B 
296 (1992) 219.

\bibitem{ill} K.F. Smith et al., Phys. Lett. B 234 (1990) 191.

\bibitem{spb} I.S. Altarev et al., Phys. Lett. B 276 (1992) 242.

\bibitem{khr} I.B. Khriplovich and S.K. Lamoreaux, CP Violation 
without Strangeness (Springer, 1998).

\end{thebibliography}
\end{document}